\documentclass[%
reprint,
superscriptaddress,
amsmath,amssymb,
twocolumn,
titlepage,
floatfix,
showpacs,
secnumarabic,
nobibnotes,
prapplied,
]{revtex4-1}

\usepackage{graphicx}
\usepackage{hyperref}
\usepackage{amssymb}
\usepackage{amsmath} 
\usepackage{textcomp}

\begin{document}

\title{Full spatiotemporal control of laser-excited periodic surface deformations}

\author{J.-E.~Pudell}
\affiliation{Institut f\"ur Physik und Astronomie, Universit\"at Potsdam, 14476 Potsdam, Germany}
\author{M.~Sander}
\affiliation{European Synchrotron Radiation Facility ESRF, 23800 Grenoble, France}
\author{R.~Bauer}
\affiliation{Institut f\"ur Nanostruktur und Festk\"orperphysik, Universit\"at Hamburg, 20355 Hamburg, Germany}
\author{M.~Bargheer}
\affiliation{Institut f\"ur Physik und Astronomie, Universit\"at Potsdam, 14476 Potsdam, Germany}
\affiliation{Helmholtz-Zentrum Berlin f\"ur Materialien und Energie GmbH, Wilhelm-Conrad-R\"ontgen Campus, BESSY II, 12489 Berlin, Germany}
\author{M.~Herzog}
\affiliation{Institut f\"ur Physik und Astronomie, Universit\"at Potsdam, 14476 Potsdam, Germany}
\email{marc.herzog@uni-potsdam.de}
\author{P.~Gaal}
\affiliation{Institut f\"ur Nanostruktur und Festk\"orperphysik, Universit\"at Hamburg, 20355 Hamburg, Germany}
\email{pgaal@physnet.uni-hamburg.de}

\date{\today}

\begin{abstract}
We demonstrate full control of acoustic and thermal periodic deformations at solid surfaces down to sub-nanosecond time scales and few-micrometer length scales via independent variation of the temporal and spatial phase of two optical transient grating (TG) excitations. For this purpose, we introduce an experimental setup that exerts control of the spatial phase of subsequent time-delayed TG excitations depending on their polarization state. Specific exemplary coherent control cases are discussed theoretically and corresponding experimental data are presented in which time-resolved x-ray reflectivity measures the spatiotemporal surface distortion of nanolayered heterostructures. Finally, we discuss examples where the application of our method may enable the control of functional material properties via tailored spatiotemporal strain fields.
\end{abstract}

\maketitle

\section{Introduction}
Ultrashort strain pulses can be generated by absorption of femtosecond or picosecond optical light pulses in solids and nanostructures \cite{Thom1986, Ruel2015}. This photoacoustic generation is employed to study properties of phonons in solids \cite{shay2013a, boja2015a} or the interaction of lattice strain with optic \cite{Sing2014}, electronic \cite{weis2014} or magnetic \cite{sche2010, Kim2012} degrees of freedom. The various interaction channels suggest that lattice strain may be used as a functional tool to control and trigger specific processes and functions in materials. Recently, the control and enhancement of quantum entanglement using tailored surface acoustic waves was suggested \cite{Blat2014a}. In fact, thanks to their customizable short length and time scales down to few nanometers and picoseconds, respectively, optically generated strain pulses may be particularly suited for selective excitation of nanostructures. However, a high level of control of the shape, frequency, lifetime, etc.\ of lattice strain is necessary before strain pulses can be used as functional tools for device operation. The control is typically gained by tailoring the temporal excitation sequence \cite{Nelson1982a,Klie2011,herz2012c,Schuelein2015}. We recently demonstrated strain control on sub-nanosecond timescales by exploiting spatial variation of transient grating excitation sequences \cite{sand2017a}. This method is not only limited to the control of coherent strain pulses but also applies to thermal deformations. In particular, the control of thermal deformations can be applied on the same timescale as the control of coherent excitations. Hence, thermal strain, which is often regarded as undesired side effect to optical excitations, can now be used to trigger specific material functions.

In this article we present a comprehensive discussion of spatiotemporal control of acoustic and thermal excitations in solids. Our method relies on shaping the temporal and spatial sequence of optical excitations using the so-called transient grating (TG) technique. In particular, our experimental TG setup allows for easy tuning of the relative spatial phase of subsequent TG excitations. The periodic surface deformation (PSD) of the sample upon optical excitation is detected using time-resolved x-ray reflectivity (\mbox{TR-XRR}) \cite{sand2017b}. The article provides a detailed discussion of spatiotemporal coherent control with particular focus on important limiting cases in section~\ref{sec:cc}. Also the quantitative probing of PSDs by diffraction of x-rays in x-ray reflection geometry is briefly explained. The optical setup for generating and controlling PSDs is presented in section~\ref{sec:setup}. In section~\ref{sec:Mod} we discuss results of spatiotemporal coherent control measurements on thermal and acoustic PSDs in nanoscopic heterostructures. We analyze and decompose the experimental data by comparison to an empirical modeling. Finally, section~\ref{sec:Conclusion} summarizes and emphasizes the main results.

\section{Coherent control of periodic surface deformations}\label{sec:cc}
Classical coherent control can be performed on any oscillator or wave-like harmonic excitation due to the superposition principle. Without restriction of the general validity, we restrict our considerations to an impulsive and displacive excitation of modes by an ultrashort excitation pulse \cite{barg2004a}, i.e., the dynamics of the involved modes is much slower than the excitation pulse duration and the coherent oscillation occurs around a displaced equilibrium. The oscillation amplitude of such an impulsively excited harmonic oscillator can either be suppressed or doubled by a second identical excitation with a relative time delay $\tau$ of a half-period (relative phase $\phi=\pi$) or a full period ($\phi=2\pi$), respectively. This type of coherent control has been successfully applied to2 various phenomena such as phonons \cite{Lindenberg2002,Synnergren2007,Beaud2007,Cheng2017}, magnons \cite{Zhang2002,Kampfrath2011,Nishitani2012}, phonon-polaritons \cite{Ward2004}, and surface acoustic waves (SAWs) \cite{Hurley2008,Li2012,sand2017a,Yang2018}. In displacive excitations, coherent control is limited to the oscillatory motion, i.e., the coherent part of the system response. The displacement from the equilibrium ground state, which for optical excitations corresponds to a heating of the sample due to the absorbed optical energy, is typically ignored in the coherent control experiments although it may contain most of the absorbed energy\cite{herz2012c}.

In addition to the temporal coherence exploited in earlier coherent control experiments, an impulsive TG excitation also possesses a spatial coherence in form of a sinusoidal intensity variation with spatial period $\Lambda$ which is typically oriented parallel to the sample surface. Assuming a linear response of the sample, the absorption of optical energy thus generates a spatially periodic energy density along the sample surface with periodicity $\Lambda$. The depth profile of the absorbed energy density is dictated by either the optical properties of the sample or the sample dimensions. In the following we will consider a thermoelastic excitation, i.e., the absorbed energy density results in a mechanical stress that eventually gives rise to a thermal transient grating (thTG). Given the impulsive excitation with ultrashort laser pulses, the thermoelastic stress also launches coherent counter-propagating surface acoustic waves (SAW) resulting in a standing SAW in the optically excited area. Both excitations can be associated with the characteristic wavevector $|\vec{q}_\parallel|=2\pi/\Lambda$. The spatial coherence introduces an additional coherent control coordinate given by the spatial phase of the TG excitation \cite{sand2017a}. Hence, a second TG excitation with a spatial phase $\varphi_x$ relative to the first TG excitation can be employed to control the relative spatial phase $\phi_\text{th}$ of the corresponding thTGs. Choosing $\phi_\text{th} = 2n \pi$ or $\phi_\text{th} = (2n-1) \pi$ either amplifies or suppresses the thTG, respectively. Here, the relative spatial phase of the TG excitation patterns directly determines the relative spatial phase of the thTGs, i.e., $\phi_\text{th}=\varphi_x$. The coherent SAW, however, is controlled via the spatiotemporal phase $\phi_\text{SAW}=\varphi_x+\varphi_t=\varphi_x+v_\text{SAW} |\vec{q}_\parallel| \tau$, where $v_\text{SAW}$ is the phase velocity of the excited SAW and $\tau$ is the temporal delay between the two TG excitation pulses. Similar to the thTG, the SAW can be amplified ($\phi_\text{SAW}=2n \pi$) or suppressed ($\phi_\text{SAW}=(2n-1) \pi$). According to the definition of $\phi_\text{SAW}$, a change in the relative spatial phase $\varphi_x$ implies an adopted time delay $\tau$ if the interference of the SAWs is to be kept unchanged. In summary, we have introduced two experimental coherent control coordinates $\varphi_x$ and $\varphi_t$ given by the relative spatial phase and the relative time delay of two consecutive TG excitations in order to control the thermal and coherent PSD via the spatiotemporal phases $\phi_\text{th}$ and $\phi_\text{SAW}$.

\begin{figure}[t]
  \centering
  \includegraphics[width = 86mm]{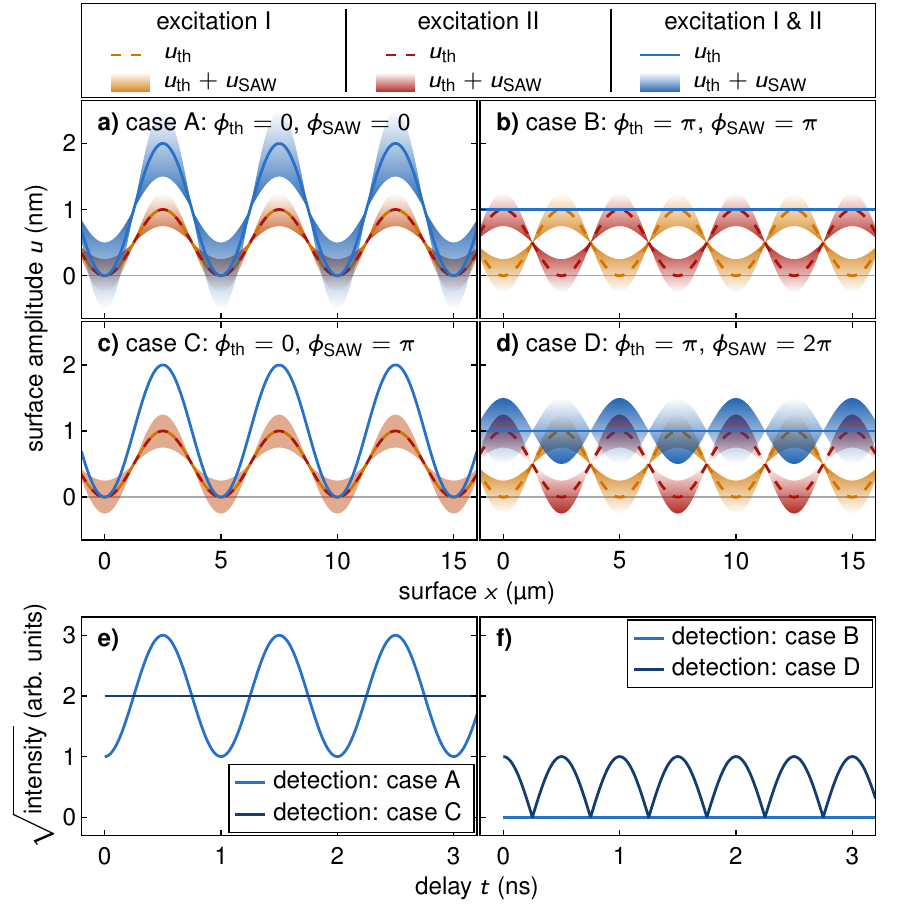}
  \caption{\textbf{Spatiotemporal coherent surface control with two tailored excitations:} yellow and red dashed lines depict quasi-static thTGs of the individual excitations I and II respectively. The gradient of the colored shaded areas represent the temporal evolution of the standing SAW on the thTG over a half-period. Blue lines and blue colored areas represent the combined excitation (excitation I \& II): a) case A: constructive interference of two thTGs and constructive interference of two standing SAWs. Note: both excitations have the same spatiotemporal phase, therefore the shaded area is orange. b) case B: destructive interference of two thTGs and destructive interference of two standing SAWs. c) case C: constructive interference of two thTGs and destructive interference of two standing SAWs. Note: both excitations have the same spatial phase and opposite temporal phase, therefore the shaded area is orange without a color gradient. d) case D: destructive interference of two thTGs and constructive interference of two standing SAWs. e) square root of the diffracted intensity from the combined excitation for case A and C; f) for case B and D. Note: for x-rays the square root of the intensity is proportional to the surface modulation amplitude.}
  \label{fig:doublefreq}
\end{figure}

In this article, we employ these coherent control coordinates to disentangle four extreme cases of coherent control, which are depicted in Fig.~\ref{fig:doublefreq}:
\begin{enumerate}
    \item[A.] Constructive interference of thTGs and constructive interference of two standing SAWs:\\
        $\varphi_x=0$, $\varphi_t = 0$ $\rightarrow$ $\phi_\text{th}= 0$, $\phi_\text{SAW} = 0$. (Fig.~\ref{fig:doublefreq}a))
    \item[B.] Destructive interference of thTGs and destructive interference of two standing SAWs:\\
        $\varphi_x=\pi$, $\varphi_t = 0$ $\rightarrow$ $\phi_\text{th}= \pi$, $\phi_\text{SAW} = \pi$. (Fig.~\ref{fig:doublefreq}b))
    \item[C.] Constructive interference of thTGs and destructive interference of two standing SAWs:\\
        $\varphi_x=0$, $\varphi_t = \pi$ $\rightarrow$ $\phi_\text{th}= 0$, $\phi_\text{SAW} = \pi$. (Fig.~\ref{fig:doublefreq}c))
    \item[D.] Destructive interference of thTGs and constructive interference of two standing SAWs:\\
        $\varphi_x=\pi$, $\varphi_t = \pi$ $\rightarrow$ $\phi_\text{th}= \pi$, $\phi_\text{SAW} = 2\pi \equiv 0$. (Fig.~\ref{fig:doublefreq}d))
\end{enumerate}
Note that while there is no SAW oscillation present in case B and C, the SAW modulates the surface deformation of a thTG or a flat surface in case A and D, respectively, as indicated by the blue shading in Fig.~\ref{fig:doublefreq}a-d). As will be discussed below, \mbox{TR-XRR} can unambiguously distinguish these situations as they will manifest differently in the experimental data. Note that we assume two identical excitations in accordance with a time-independent thTG. The amplitude of the second TG excitation may of course be adapted to compensate for a potential decay of the first excitation during the time delay $\tau$.

In the following we briefly discuss the detection of dynamics triggered by the TG excitation. Typically, experiments employ an optical laser pulse which is diffracted from the excited modes via photothermal or photoelastic effects \cite{Mazn2003, Vega2015}. Alternatively, PSDs can be detected using x-ray diffraction and photoemission electron microscopy techniques \cite{Sauer1999,Nicolas2014,Foerster2019}. We have recently shown that the PSD associated with the excited quasi-static and transient modes may also be probed by \mbox{TR-XRR} \cite{sand2017a,sand2017b}. This method is exclusively sensitive to the surface displacement and can detect deformations of only few nanometers \cite{sand2017a}. Thermally induced expansion can be unambiguously disentangled from coherent elastic effects via the characteristic timescale of the surface deformation \cite{sand2017b}, thus yielding a complete picture of the surface dynamics. For the \mbox{TR-XRR} probe the Laue condition, i.e., momentum conservation, must be fulfilled
\begin{equation}
    \vec{k}_{\pm1}=\vec{k}_\text{in}+\vec{q}_{\perp}\pm\vec{q}_{\parallel} \label{equ:Laue}
\end{equation}
where $\vec{k}_\text{in}$ and $\vec{q}_{\perp}$ are the wavevector of the incident probe photons and the recoil momentum due to reflection at the surface, respectively. We restrict our considerations to first-order diffraction from the PSD although diffraction to higher orders is also present even if a perfect sinusoidal PSD is monitored \cite{sand2017b}. Note that Eq.~\eqref{equ:Laue} is independent of the spatial phase of the TG, i.e., the phase of the PSD cannot be inferred from the diffracted intensity per se. The intensity of the diffracted x-ray probe pulse is proportional to $\Delta u^2(t)$, i.e., to the squared difference between the minimum and maximum surface deformation\cite{sand2017b}. Thus, the square root of the diffracted intensity measures the magnitude of the total surface modulation amplitude, i.e., $\sqrt{I(t)}=|\Delta u(t)|= |\Delta u_\text{th}+\Delta u_\text{SAW} \cos(\omega t)|$, where $\omega=v_\text{SAW}|q_{\parallel}|$ is the frequency of the SAW.

First consider the case where a strong thTG is modulated by a SAW with relatively small amplitude, i.e., $\Delta u_\text{SAW} < \Delta u_\text{th}$. Excluding thermal diffusion within the thTG, $\Delta u(t)>0$ holds for all times and the measured signal directly reveals the SAW's frequency and its relative phase with respect to the thTG as depicted in Fig.~\ref{fig:doublefreq}e). However, if $\Delta u_\text{SAW} > \Delta u_\text{th}$ the detection signal is altered. The extreme case is shown in Fig.~\ref{fig:doublefreq}d), where the SAW modulates a flat surface (case D), i.e., $\Delta u_\text{th}=0$ (note that $u_\text{th}\neq 0$ may hold). Here, the time-dependent intensity is proportional to $\Delta u^2(t)=\left[\Delta u_\text{SAW} \cos(\omega t)\right]^2 = \Delta u_\text{SAW}^2 \left[ 1+ \cos(2\omega t)\right]/2$. Therefore, the diffracted probe intensity shows twice the frequency of the SAW as presented in Fig.~\ref{fig:doublefreq}f). The effect of the thTG is analogous to a spatial local oscillator, which allows inferring the spatiotemporal phase of the coherent SAW from the diffracted probe pulse after TG excitation.

\section{Experimental Methods}\label{sec:setup}
\begin{figure*}
  \centering
  \includegraphics[width = \textwidth]{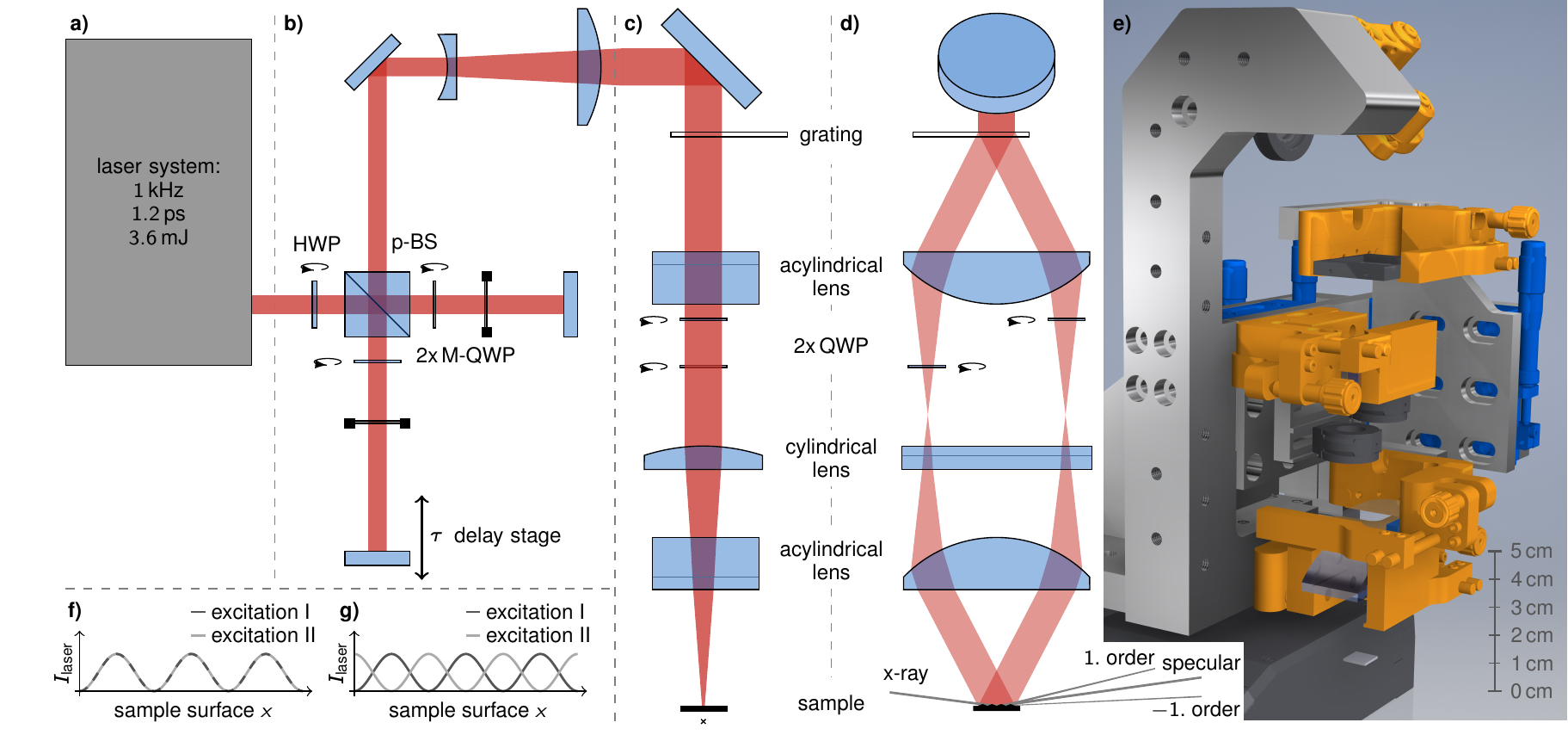}
  \caption{\textbf{Optical setup:} a) The output of a 1\,kHz laser system is coupled to a Michelson interferometer [b)], where s- and p-polarized replica of the pump pulse are generated. c) shows a side view of the TG-setup. The front view d) shows the beam path of the $+1$st and $-1$st diffraction order. A 3D-view d) of the TG-setup shows the opto-mechanical components. In f) and g) the created relative spatial phase of the interference grating for parallel and perpendicular setting of the QWPs are shown.}
  \label{fig:setup}
\end{figure*}
A detailed view of our optical experimental setup is shown in Fig~\ref{fig:setup}a-e). The general layout of the TG setup is described elsewhere\cite{sand2017a, Janu2016a}. Our particular setup was designed for optical \mbox{pump--TR-XRR} probe measurements at the ID09 beamline at the European Synchrotron Radiation Facility (ESRF). Taking into account the specific conditions at the beamline, we optimized the optical setup for small size, stability and tunability. Therefore, we use two 50\,mm wide acylindrical lenses with a focal length of 40\,mm to image the $+1$st and $-1$st diffraction order from a transmission phase mask onto the sample surface. Interference of the +1 and $-$1 beam at the sample surface results in a sinusoidal modulation of the optical intensity. Grating, lenses and sample were mounted in 4f-geometry. We optionally introduce an additional cylindrical lens with a focal length of 75.6\,mm in the perpendicular plane. This lens is mounted with variable distance to the sample to generate higher excitation fluences. 

To generate two replica of an ultrashort optical pump pulse with perpendicular polarization for coherent control we use a Michelson interferometer [cf.~Fig.~\ref{fig:setup}b)] with a polarizing beamsplitter (PBS). The relative intensity of the pulses can be tuned by a half-wave plate (HWP) in front of the PBS. Each arm of the interferometer includes a quarter-wave plate (\mbox{M-QWP}) which, due to the double-passage of the beams, effectively rotates the linear polarization by 90$^{\circ}$. Thus, the output of the interferometer yields one s- and one \mbox{p-pol}arized optical pump pulse. The pulses have a variable relative time delay $\tau$ that is defined by the difference of the path lengths of the Michelson interferometer arms. Both pulses are subsequently coupled into the TG setup where they essentially are diffracted into $\pm$1st order beams by the phase mask. Zero-order and higher-order diffraction intensities are minimized by the specific design of the transmission phase mask.

In addition to simple TG excitation, our setup allows for selecting the spatial phase of the TG to perform spatiotemporal coherent control. We use the fact that the spatial phase of the TG at the sample surface depends on the relative temporal phase of the interfering optical beams. For example, interference gratings generated by two optical beams of either parallel or anti-parallel polarization have opposite spatial phase with respect to one another\cite{Laba1999a}. Note that in the anti-parallel case the electric fields of $+1$st and $-1$st order beams have a relative temporal phase of $\pi$, i.e., exactly the same value as the relative spatial phase shift of the generated TG. As explained below, we employ quarter-wave plates (QWP) and different polarizations to impose a relative spatial phase shift between two consecutive TG excitations.

After collimation by the first acylindrical lens, each beam propagates through a QWP. The QWPs are oriented either with the fast or slow axis aligned with the polarization of the laser pulses. If the QWPs have identical orientation, there is no relative temporal phase offset between the $\pm$1st order beams as they both traverse the QWPs at either the fast or the slow axis. Note that both consecutive s- and p-polarized laser pulses generate TGs with identical spatial phase, i.e.\ $\varphi_\text{x}=0$. Hence, with a parallel setting of the QWPs, one can generate cases A and C discussed in section~\ref{sec:cc}.

If the QWPs are oriented perpendicular to each other, the $+1$st and $-1$st order beam experience a relative temporal phase shift of $\pm\pi/2$ which directly translates into a spatial phase offset of $\pm\pi/2$ of the TG excitations. The opposite sign of the temporal phase shift holds for the s- and p-polarized beams, respectively. The magnitude of the spatial phase difference between the s- and p-polarized TG is therefore equal to $\pi$, i.e.\ $\varphi_\text{x}=\pi$. Hence, with a perpendicular setting of the QWPs, one readily obtains cases B and D discussed in section~\ref{sec:cc} where the PSD due to the thTG is relieved by the second TG excitation.

The experimental results presented and discussed in the next section have been obtained on a 30\,nm thick metallic SrRuO$_3$ (SRO) film epitaxially grown by pulsed laser deposition on a (110)-oriented DyScO$_3$ (DSO) substrate. The sample was excited with TG excitations each having a spatial period $\Lambda=2.4$\,\textmu m and an incident pump fluence of 18\,mJ/cm$^{2}$ for the central fringes of the TG\footnote{The incident laser power per TG excitation was 970\,mW at 1\,kHz repetition rate. The footprint of the non-interfering laser beams at the sample surface (elliptical Gaussian) was $4.0\times3.4$\,mm$^2$ (major and minor 1/$e$-diameter). Assuming a homogeneous distribution of the laser pulse energy over an ellipse having the above size yields a pump fluence of 9\,mJ/cm$^{2}$. Due to the interference of the crossing laser beams the TG fringes at the center of the elliptical Gaussian will have twice that fluence in the maximum}. We employ a commercial Ti:Sapphire laser amplifier (Coherent Legend Elite) which delivers 800\,nm pulses with a duration of 1.2\,ps and a pulse energy of 3.6\,mJ. The laser repetition rate is 1\,kHz synchronized to the synchrotron. The shortest grating period $\Lambda$ inscribed in the sample is ultimately limited by the laser wavelength. By frequency doubling or tripling of the fundamental frequency one can reduce the period to less then 300\,nm. The generation of transient gratings with periods less than 100\,nm has been demonstrated by using high-energy radiation from free electron laser sources \cite{Mazn2018a,Svet2019a}. Thus, our method can be employed truly on nanometer length scales.

Monochromatized 15\,keV x-ray probe pulses were selected from the synchrotron pulse train by a high-speed chopper at the same frequency. The pump-probe delay can be changed electronically by the laser synchronization unit. In the experiment presented here, the total temporal resolution was limited to 75\,ps mainly due to the duration of the x-ray probe pulses. In principle, the experimental time resolution is also limited by the rather large wavefront tilt between exciting laser and probing x-ray pulses, however, in the present case this is only a minor limitation (approx.\ 10\,ps). Diffracted x-ray photons where captured on an area detector (Rayonics \mbox{MX170-HS}).\cite{wulf2003,camm2009} For the evaluation the intensity $I_\text{$-1$st}$ of the $-1$st diffraction order is integrated in a region of interest on the area detector. The recorded intensity is normalized to a static diffraction background $I_\text{n}$ to reduce influences of beam instabilities and thermal drifts of sample and setup. In order to extract the surface modulation amplitude $\Delta u$, we take the square root of the diffracted intensity after subtracting a scattering background $I_\text{bg}$ by averaging all unpumped detected intensities of the $-1$st diffraction order. This results in 
\begin{equation}
    \Delta u \propto \sqrt{\left|\frac{I_\text{$-1$st} -  I_\text{bg}}{I_\text{n}}\right|} \text{sgn}\left(I_\text{$-1$st} -  I_\text{bg}\right)
\end{equation}
where the absolute function circumvents imaginary results and sgn function projects these values on the negative axis for the surface modulation amplitude $\Delta u$.

\section{Results and Discussion}\label{sec:Mod}
\begin{figure}
    \centering
    \includegraphics[width = 86mm]{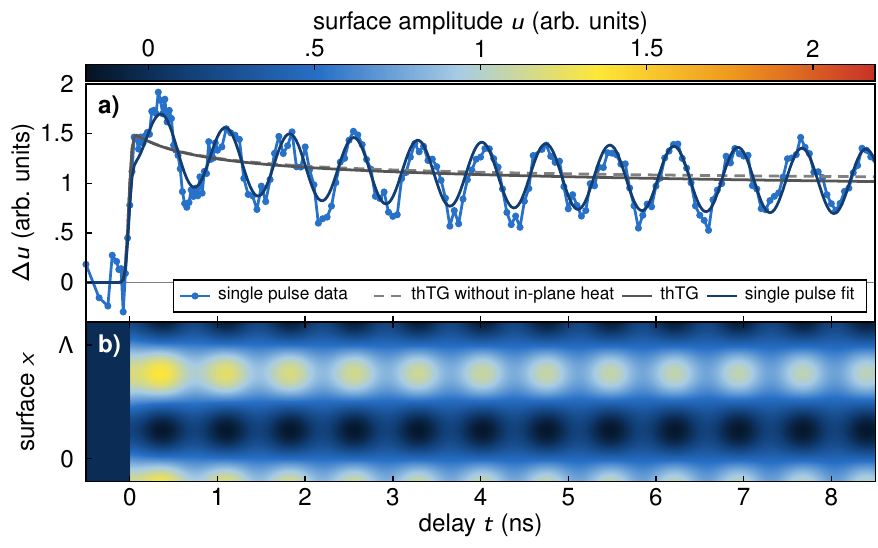}
    \caption{\textbf{Single-pulse excitation:} a) Time-dependent surface modulation amplitude $\Delta u(t)$ measured with time-resolved x-ray reflectivity (\mbox{TR-XRR}). The data (blue bullets) is modeled by calculating the spatiotemporal surface deformation field $u(x,t)$ [c.f.\ Eq.~\eqref{eq:mod}]. The dark blue solid line shows the time dependence of the Fourier component $(\mathcal{F}u)(q_\parallel,t)$ at the characteristic wavevector $q_\parallel$. The surface modulation amplitude of only the thermal distortion $(\mathcal{F}u_\text{th})(q_\parallel,t)$ due to the thTG is shown by the gray lines with (solid) and without (dashed) in-plane heat diffusion. b) Illustration of the spatiotemporal surface deformation field $u(x,t)$ calculated using Eq.~\eqref{eq:mod}.}
    \label{fig:sp}
\end{figure}

First, we briefly discuss the transient response of the sample surface to a single TG excitation. As derived in earlier studies \cite{sand2017b}, the surface modulation amplitude $\Delta u$ is proportional to the square root of the diffracted intensity. The transient amplitude of the laser-generated PSD inferred from the the x-ray intensity diffracted into $-1$st order is shown by the blue bullets in Fig.~\ref{fig:sp}a).
It features a step-like rise followed by oscillations on top of a slowly decaying thermal offset. The surface excursion field can thus be precisely modeled by
\begin{equation}
    u = \Theta(t)\Big[ u_\text{th}(x,t,\varphi_\text{x}) + u_\text{SAW}(x,t,\varphi_\text{x},\varphi_\text{t})\Big] \label{eq:mod}
\end{equation}
where $u_\text{th}$ is the slowly decaying amplitude of the thTG and $u_\text{SAW}$ is the amplitude of the coherent surface acoustic mode defined as
\begin{eqnarray}
    u_\text{th} &=& \frac{u_\text{th,0}(t)}2\, e^{-\alpha_x q_\parallel^2 t} \left[1+\sin\left(q_\parallel x-\varphi_\text{x}\right)\right] \label{eq:tg} \\
    u_\text{SAW} &=& -\frac{u_\text{SAW,0}}2\, \sin\left(q_\parallel x-\varphi_\text{x}\right)\cos(\omega t-\varphi_\text{t}) \label{eq:saw}
\end{eqnarray}
The rise time $\tau_\text{rise}$ of the thTG is dictated by the ratio of the thickness (or the optical penetration depth if the latter is much shorter) and sound velocity of the laser-excited film. Typical time scales of thin-film expansion are a few tens of picoseconds or even down to a few picoseconds for very thin films \cite{Schick2014b}. In the present case $\tau_\text{rise}$ is much shorter than all other involved dynamics and thus approximated by the Heaviside function $\Theta(t)$ in Eq.~\ref{eq:mod}. In fact, $\tau_\text{rise}$ defines the fundamental limit for coherent control of the thTG (case B and D) which can thus be truly applied down to picosecond time scales as demonstrated in Ref.~\citenum{sand2017b}. The concept of spatiotemporal coherent control is generally applicable to an arbitrary number of coherent modes \cite{sand2017a,sand2017b,chan2017}, but in the present case the data only exhibit a single Rayleigh-like SAW mode. We can thus restrict our model to only include this single coherent mode. Note that the first TG excitation always defines the zero phase $\varphi_x$ and $\varphi_t$, respectively. In order to mimic the sensitivity of the x-ray probe beam to only the modulation of the PSD, we extract the transient wavevector-dependent surface modulation amplitude $\Delta u(q,t) = (\mathcal{F}u)(q,t)$ by Fourier transformation of the spatiotemporal surface deformation field $u(x,t)$ depicted in Fig.~\ref{fig:sp}b). We than evaluate the surface modulation amplitude $\Delta u(q_\parallel, t)$ at the characteristic wavevector $q_\parallel$. The dark blue solid line in Fig.~\ref{fig:sp}a) shows the temporal behavior convoluted with the experimental temporal resolution of 75\,ps.

\begin{figure*}[t]
    \centering
    \includegraphics[width = \textwidth]{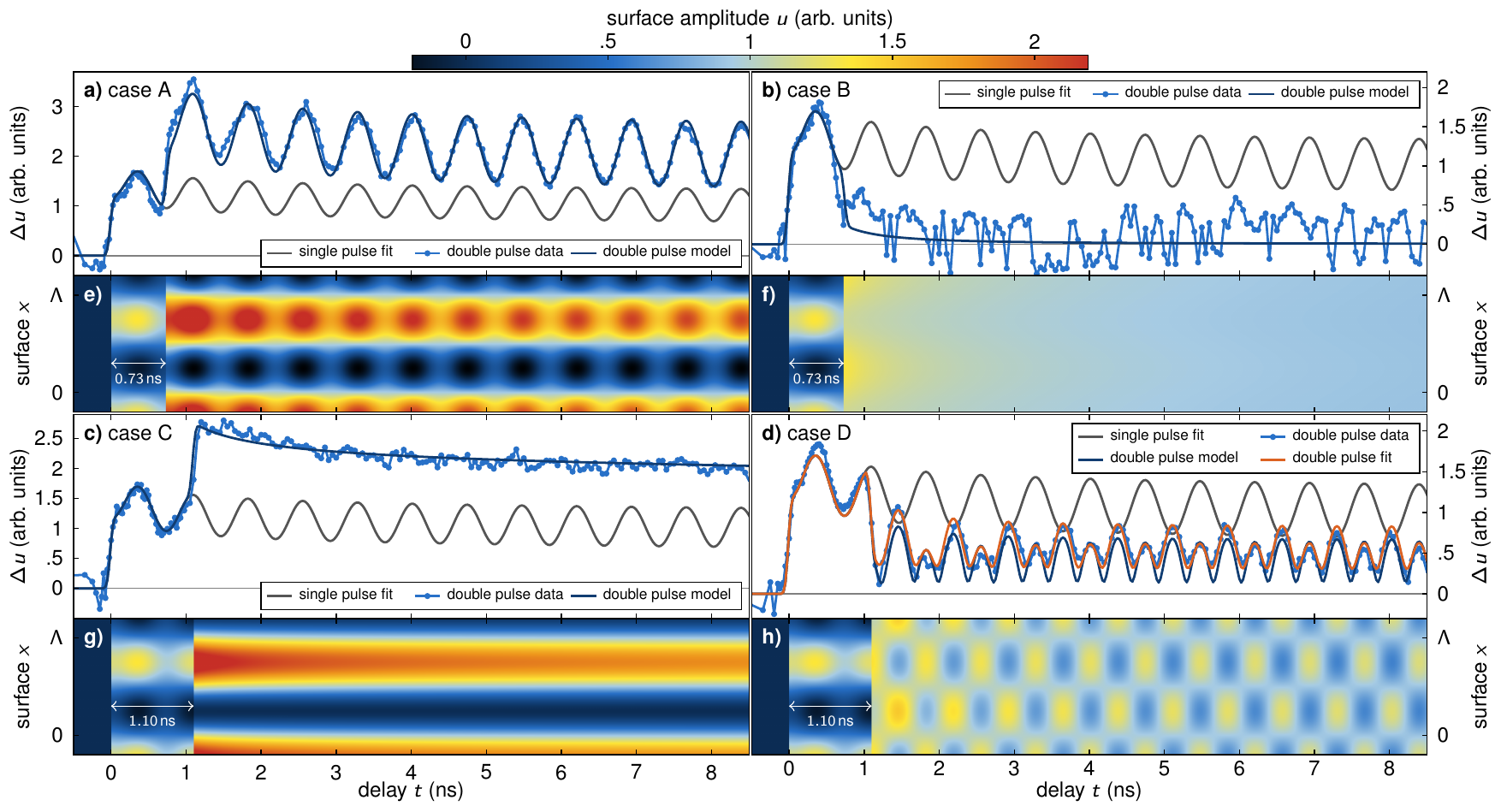}
    \caption{\textbf{Double-pulse excitation:} a,e) case A; b,f) case B; c,g) case C; d,h) case D. The single pulse response as modeled in Fig.~\ref{fig:sp} are shown in gray for comparison in a-d). Symbols show the TR-XRR measurements and dark blue solid lines show the transient surface modulation amplitude $\Delta u(q_\parallel,t)$ for the characteristic wavevector $q_\parallel$ of the spatiotemporal surface deformation field shown in e-h) derived from Eq.~\eqref{eq:mod}. All calculated traces are convoluted with a 75\,ps Gaussian to match the temporal resolution of the experiment. The orange line in d) represents the best fit to the double pulse data by allowing for an amplitude and phase variation of the second excitation pulses.}
    \label{fig:cc}
\end{figure*}
The slowly decaying signal shown in Fig.~\ref{fig:sp}a) is caused by the PSD associated with the thTG that slowly decays due to thermal diffusion. This decay clearly exhibits a fast component decaying within the first 2\,ns and a much slower component. The latter is due to in-plane thermal diffusion between hot and cold areas of the thTG. An analytical solution of the in-plane thermal diffusion for a sinusoidal thermal grating yields the exponential term in Eq.~\eqref{eq:tg} which implies a decay time of $(\alpha_{x}q_\parallel^{2})^{-1}=210$\,ns \cite{Kaed1995,Johnson2012}. Here, the homogeneous in-plane diffusivity $\alpha_{x}=0.8$\,mm$^2$/s is used for a substrate temperature of 323\,K \cite{HIDDE2018}. The initial fast decay originates from different thermal expansion coefficients of the metallic SRO film and the insulating DSO substrate. In fact, SRO expands stronger upon heating than DSO \cite{Yamanaka2004,UECKER2008}. Thus, as heat diffuses along the out-of-plane direction from the excited areas in the SRO film into the substrate, the total surface excursion is reduced. In order to verify this, we model the out-of-plane heat transport by solving the one-dimensional heat diffusion equation with a finite-element method \cite{Shayduk2011,schi2014} by accounting for all relevant thermophysical properties of the materials. The experimental data is reproduced without including additional thermal resistance at the interface due to the nearly perfect acoustic impedance match between the involved materials and the high structural quality of the sample. The simulation yields the surface distortion $u_\text{th,0}(t)$ used as input for Eq.~\eqref{eq:tg}. The surface modulation amplitude due to combined out-of-plane and in-plane thermal diffusion is shown as a gray solid line in Fig.~\ref{fig:sp}a). If in-plane thermal diffusion is neglected $(\alpha_x=0)$, the grey dashed line is obtained which proves that the initial fast decay is indeed governed by the out-of-plane thermal transport. The very good agreement between experiment and calculation evidences that the presented method can be a powerful tool to investigate multidirectional thermal transport in nanoscopic heterostructures. We do not observe deviations from linear behaviour of our sample even up to very large excitation densities \cite{herz2012a}. This aspect is particularly important for the multipulse excitation discussed in the next paragraph.

With the well-calibrated single-pulse excitation we finally demonstrate full spatiotemporal control of transient and quasi-static PSDs via double-pulse TG excitation. By employing two consecutive TG excitations, we set the spatiotemporal phase of the excitation to the four different cases introduced in Sec.~\ref{sec:cc}. The corresponding dynamic surface excursion measured by \mbox{TR-XRR} for these four cases is shown in Fig.~\ref{fig:cc}a-d), respectively. As predicted above, the spatial phase setting $\varphi_x=0$, selected by a parallel alignment of the QWPs in the optical setup, results in an increase of the PSD (case A and C) due to an enhancement of the thTG. In contrast, the 90$^\circ$ rotation of only one QWP suppresses the PSD and relieves the thTG completely (case B and D). The phase of the SAW is controlled by both the spatial and temporal phase $\varphi_x$ and $\varphi_t$, respectively. For any setting of the spatial phase, one can either suppress (case C and B) or enhance (case A and D) the SAW mode by choosing the correct time delay $\tau$ of the second TG excitation. Note in particular the comparison of cases A and B, where $\tau=0.73$\,ns is identical. Still, not only the timing $\tau$ alone determines the amplitude of the SAW after the second excitation, as explained in Sec.~\ref{sec:cc}.

The possibility of suppressing the coherent mode (case B and C) offers a precise tool for investigations of the multidirectional thermal transport in nanoscale heterostructures without undesired coherent signals yet ensuring sufficient time resolution given by the ultrashort laser and x-ray pulses. In case C we clearly observe the multicomponent relaxation due to in-plane and out-of-plane heat diffusion which was discussed above for single-pulse excitation. However, here, the signal of thermal origin is not masked by the coherent signal. The finite decaying intensity after the second TG excitation in case B (Fig.~\ref{fig:cc}b)) evidences that the thTG was not immediately suppressed by the second excitation. This is caused by the partial decay of the first thTG between the two excitations due to fast out-of-plane heat diffusion which results in the observed imbalance of both thTGs. All observations are accurately reproduced by our modeling introduced above. Altogether, the sensitivity to thermal transport in layered heterostructures can be greatly enhanced with spatiotemporal coherent control using TG excitations.

A qualitative difference from the other recorded signals is observed in case D shown in Fig.~\ref{fig:cc}d) where the thTG is suppressed and the SAW is enhanced. Here, we observe a coherent oscillation exhibiting twice the frequency of the excited SAW. Recall that the data represent the variations of the 1st-order diffracted x-ray intensity. If a true second harmonic of the fundamental SAW was present, Eq.~\eqref{equ:Laue} implies that the corresponding 1st order of the second harmonic would be diffracted towards larger angles. In other words, the 1st-order diffraction angle cannot contain signatures of a second harmonic SAW. Again, we model the transient diffracted x-ray intensity caused by the spatiotemporal sample surface dynamics in analogy to the single-pulse excitation data analysis [c.f.\ Eqs.~(\ref{eq:mod}-\ref{eq:saw})]. Note that our model is a purely linear response model and thus does not support higher harmonics of the employed modes. Assuming the time delay $\tau$ chosen in the experiment and two identical TG excitations we obtain the dark blue curve in Fig.~\ref{fig:cc}d). Obviously, our modeling correctly yields the doubled frequency. As indicated above this is due to the detection process which measures different oscillation frequencies from the same acoustic mode with and without the additional thTG that acts as a spatial local oscillator. Note that the amplitudes of the even and odd oscillation maxima are different which can be traced back to the heat diffusion dynamics of the first thTG between the two TG excitations breaking the symmetry. As case B revealed, the two thTGs converge in amplitude after a few nanoseconds. In case D (dark blue line) this is manifested in the equilibration of the even and odd oscillation maxima on the same timescale. However, this equilibration of oscillation amplitudes is not fully featured in the experimental data which indicates an asymmetry in the TG excitation strengths. Also, the level around which the signal oscillates is larger. Indeed, the data are accurately reproduced (orange line in Fig.~\ref{fig:cc}d)) if we assume a 9\% larger amplitude and a slight detuning of the spatial phase of 3\% for the second excitation pulse. Such errors may stem from slight deviations from optimal laser beam and QWP alignments.

\section{strain control in functional materials}
In the following paragraph we outline possible interaction channels between the excited deformation and functional properties of the crystal. In particular, we describe strain-induced changes of the free energy density via magnetoelastic effects and changes of the electronic band energy via deformation potential coupling. Finally, we will briefly introduce active optical elements which use dynamic strain fields to manipulate x-ray pulses emitted by synchrotron storage rings.

If strain is used as a functional tool, it is important to recall, that the interaction of lattice deformations with a material strongly depends on the symmetry, i.e., on the specific component of the strain tensor $\varepsilon$. The strain fields corresponding to thTG and coherent Rayleigh-like SAWs are composed of both compressive/tensile (e.g., $\varepsilon_{xx}$, $\varepsilon_{zz}$) and shear (e.g., $\varepsilon_{xz}$) components. The \mbox{TR-XRR} method detects the absolute surface deformation, i.e., the integrated out-of-plane expansion of the excited volume. However, knowing one component of the strain tensor of a Rayleigh wave allows inferring all other components as well.\cite{Hess2002} In our coherent control scheme, all strain components of the coherent mode are customizable as well as the in-plane and shear components of the thTG. Only the out-of-plane component of the thermal grating is given by the initially absorbed energy density profile.

In multiferroic materials a dynamic strain wave modifies the free energy density due to elastic deformations of the lattice. As an example, we discuss ferromagnetic materials, where the magnetoelastic interaction modulates the free magnetic energy density $f_\text{mag}$.\cite{sand1999a} In a static case, $f_\text{mag}$ is composed of the Zeeman energy, which depends on an external magnetic field, and of static anisotropy components such as magneto-crystalline, shape and magnetoelastic anisotropy.\cite{Hand2000} The interplay of these terms results in a direction and magnitude of the macroscopic magnetization $\vec{M}$. Their dynamics can be induced through time-dependent changes of free magnetic energy $f_\text{mag}$. Prominent examples are ferromagnetic resonance (FMR) measurements\cite{Farl1998}, which act on the Zeeman energy or all-optical switching\cite{Kirilyuk2010}, where laser-induced heating leads to changes of the shape anisotropy. In complete analogy, an acoustic waves dynamically changes $f_\text{mag}$ via the magnetoelastic energy term\cite{sand1999a}.

Although magnetoelastic interaction is well-known, magnetoacoustics has only been investigated quite recently \cite{Kim2012,Deb2018,Zeus2019}. Since then, strain-induced magentization dynamics of nanoparticles excited specifically by Rayleigh waves, has gained strong interesst \cite{yaha2014a,teja2017a}. These efforts are driven by the potential of strain-induced dynamics, i.e., energy efficiency, mode selectivity and the ability to tailor the excitation to nanosize dimensions. Several of these recent experiments use optical generation of strain waves, thus pushing magnetoelastic excitations to picosecond timescales \cite{Janu2016a,chan2017,chang2018}. The strain control scheme described in this article not only allows to selectively excite magnetization dynamics, but also enables control of theses excitations on picosecond timescales. In particular, this is not only limited to the coherent strain but rather extends to thermal strain, while maintaining the high temporal resolution.

The second interaction channel we discuss is deformation potential coupling of electrons with acoustic phonons. The deformation of the crystal lattice by an acoustic lattice distortion leads to an energy shift $\Delta E = a \varepsilon$ of the extremal points of the electron bands where $a$ is the deformation potential which typically has a value of about 10\,eV at the $\Gamma$-point of tetrahedral semiconductors such as Si or GaAs.\cite{Blach1984a} Hence, already a small dynamic strain of the order of $10^{-3}$ up to $10^{-2}$ leads to changes of the conduction and valence bands of 10 to 100\,meV. Strain-induced changes of the electronic structure affect charge transport and optical properties \cite{Wang2016a,Cout2009} and allow for control of recombination dynamics in nanostructures \cite{weis2014,Lima2005}. Strain-control of optical properties of nanostructures is a promising candidate for applications in quantum computation and quantum information technology \cite{Barn2000,Schu2015}. The realization of such applications depends on the ability to control the lattice strain, ideally on short to ultrashort timescales. While the two examples given above may require probing mechanisms other than TR-XRR (e.g.\ magnetooptical probing, optical and/or x-ray dichroism, or valence spectroscopies), the method presented in this paper may pave the way for these future applications.

Finally, we discuss a specific application developed by our group, where strain-induced deformations are used to realize active ultrafast x-ray optics. The devices are optimized for installation at synchrotron beamlines. A prominent example is the picosecond Bragg switch (PicoSwitch), which shortens an incident synchrotron x-ray pulse to a duration of few picoseconds \cite{sand2018}. The coherent control of thTGs similar to case B [c.f. section~\ref{sec:cc}] allows for controlling diffraction of an incident x-ray pulse into the $\pm$1st diffraction order of the thTG. In particular, our approach allows to turn the diffraction on and off on sub-nanosecond timescales. Thus, thTGs could be employed to pick individual x-ray pulses from a synchrotron pulse train for subsequent pump-probe experiments. Furthermore, the device may also be employed as variable beam splitter in order to, e.g., distribute x-ray pulses among multiple beamlines. This may be particularly interesting at x-ray free electron laser (XFEL) facilities, where currently only one experimental station is operational at a time. With such an approach several instruments could be supplied with XFEL pulses in parallel. The main challenge for this device is to achieve high diffraction efficiencies. Our previous studies suggest that a maximum efficiency of more than 30\% could be reached \cite{sand2017a, sand2017b, Vadi2017}.

\section{Conclusion}\label{sec:Conclusion}
In conclusion, we demonstrated spatiotemporal control of acoustic and thermal deformations of solid surfaces. The optical setup allows for generation of transient surface gratings with variable spatial phase. Hence, a thermal deformation can either be enhanced or suppressed by a temporal sequence of excitation pulses on timescales much shorter than the deformation lifetime. In addition, we showed that the suppression of the coherent signal facilitates investigations of multidirectional thermal transport in nanolayered heterostructures with high time resolution. 
We believe that our method presents an important step towards developing strain as functional tool for solids and nanostructures. As examples we discuss the magnetoelastic interaction in ferromagnetic materials. While numerous recent studies have demonstrated the ability to manipulate the macroscopic magnetization with coherent strain pulses, our new scheme paves the way for controlled strain-induced preparation of a ferromagnetic state. Strain control may also be applied to manipulate electronic states in bulk and low-dimensional semiconductors. Finally, we discuss active optical elements, which are a new kind of strain-based devices for ultrafast x-ray beam manipulation at synchrotrons.

\section*{Acknowledgements}\label{sec:Ackn}
The TR-XRR experiments were performed at the beamline ID09 of the European Synchrotron Radiation Facility (ESRF), Grenoble, France. We are grateful to \mbox{Michael} \mbox{Wulff} and \mbox{Norman} \mbox{Kretzschmar} for providing assistance in using beamline ID09. We also gratefully acknowledge technical support of \mbox{Christine} \mbox{Fischer} and \mbox{Elko} \mbox{Hannemann}. Finally, we thank \mbox{Jutta} \mbox{Schwarzkopf} from Leibniz-Institut f\"ur Kristallz\"uchtung, Berlin for providing the sample. We acknowledge the Deutsche Forschungsgemeinschaft for the financial support via BA2281/8-1 and funding from the BMBF via FK05K16GU3.

\end{document}